\documentclass[%
superscriptaddress,
amsmath,
amssymb,
aps,
physrev,
twocolumn,
]{revtex4-2}

\usepackage[]{graphicx}
\usepackage{dcolumn}
\usepackage{bm}
\usepackage[colorlinks=true, allcolors=blue]{hyperref}
\usepackage{booktabs}
\usepackage{cancel}
\usepackage{braket}
\usepackage{multirow, tabularx}
\usepackage{amsmath, amsthm, amssymb}
\usepackage{comment}
\usepackage[toc,page]{appendix}
\usepackage[dvipsnames]{xcolor}
\usepackage{svg}

\usepackage{dsfont}
\usepackage{listings}
\usepackage{tabularray}
\UseTblrLibrary{booktabs}
\usepackage{siunitx}

\newcommand{\taumod}{\tau_{\rm mod}}
\newcommand{\taud}{\tau_{\rm d}}
\newcommand{\taus}{\tau_{\rm s}}
\newcommand{\tauop}{\tau_{\rm op}}
\newcommand{\dd}{{\rm d}}
\newcommand{\Hsagnac}{H_{\rm Sagnac}}
\newcommand{\hsagnac}{h_{\rm Sagnac}}
\newcommand{\halfwavev}{V_{\frac{\pi}{2}}}
\newcommand{\tsym}{t_{\text{sym}}}

\begin{document}

\title{General model and modulation strategies for Sagnac-based encoders}

\author{Federico~Berra}
\affiliation{Dipartimento di Ingegneria dell'Informazione, Universit\`a degli Studi di Padova, Via Gradenigo 6B - 35131 Padova, Italy.}
\author{Matías~Rubén~Bolaños}
\affiliation{Dipartimento di Ingegneria dell'Informazione, Universit\`a degli Studi di Padova, Via Gradenigo 6B - 35131 Padova, Italy.}
\author{Alberto~De~Toni}
\affiliation{Dipartimento di Ingegneria dell'Informazione, Universit\`a degli Studi di Padova, Via Gradenigo 6B - 35131 Padova, Italy.}
\author{Kannan~Vijayadharan}
\affiliation{Dipartimento di Ingegneria dell'Informazione, Universit\`a degli Studi di Padova, Via Gradenigo 6B - 35131 Padova, Italy.}
\author{Costantino~Agnesi}
\affiliation{Dipartimento di Ingegneria dell'Informazione, Universit\`a degli Studi di Padova, Via Gradenigo 6B - 35131 Padova, Italy.}
\affiliation{Padua Quantum Technologies Research Center, Universit\`a degli Studi di Padova, via Gradenigo 6B, IT-35131 Padova, Italy}
\author{Marco~Avesani}
\affiliation{Dipartimento di Ingegneria dell'Informazione, Universit\`a degli Studi di Padova, Via Gradenigo 6B - 35131 Padova, Italy.}
\affiliation{Padua Quantum Technologies Research Center, Universit\`a degli Studi di Padova, via Gradenigo 6B, IT-35131 Padova, Italy}
\author{Andrea~Stanco}
\affiliation{Dipartimento di Ingegneria dell'Informazione, Universit\`a degli Studi di Padova, Via Gradenigo 6B - 35131 Padova, Italy.}
\affiliation{Padua Quantum Technologies Research Center, Universit\`a degli Studi di Padova, via Gradenigo 6B, IT-35131 Padova, Italy}
\author{Paolo~Villoresi}
\affiliation{Dipartimento di Ingegneria dell'Informazione, Universit\`a degli Studi di Padova, Via Gradenigo 6B - 35131 Padova, Italy.}
\affiliation{Padua Quantum Technologies Research Center, Universit\`a degli Studi di Padova, via Gradenigo 6B, IT-35131 Padova, Italy}
\author{Giuseppe~Vallone}
\affiliation{Dipartimento di Ingegneria dell'Informazione, Universit\`a degli Studi di Padova, Via Gradenigo 6B - 35131 Padova, Italy.}
\affiliation{Padua Quantum Technologies Research Center, Universit\`a degli Studi di Padova, via Gradenigo 6B, IT-35131 Padova, Italy}

\begin{abstract}

{In recent decades, there has been an increasing demand for faster modulation schemes. 
Electro-optic modulators are essential components in modern photonic systems, enabling high-speed control of light for applications ranging from telecommunications to quantum communication. 
Conventional inline and Mach-Zehnder modulators, while widely adopted, are limited by bias drift, high operating voltages, and polarization-mode dispersion.
Sagnac loop-based modulators have recently emerged as a promising alternative, offering inherent stability against environmental fluctuations and eliminating the need for active bias control. 
In this work, we present a comprehensive model of the Sagnac modulator that captures both intensity and polarization modulation. 
We analyze the role of asymmetry in the loop, highlighting its impact on the achievable repetition rate, and propose modulation strategies to overcome these constraints. 
Finally, we investigate the symmetric Sagnac configuration and demonstrate practical techniques for achieving robust modulation while mitigating experimental challenges. 
Our results establish the Sagnac modulator as a versatile and stable platform for next-generation photonic and quantum communication systems.}

\end{abstract}

\maketitle

\section{Introduction}

Electro-optic modulators (EOMs) have emerged as indispensable components of photonic systems, finding widespread applications in metrology \cite{carlson2018ultrafast}, data processing, and high-capacity optical communications \cite{Xu2020highperformance}, due to their capability for high-speed, high-bandwidth  operation \cite{wooten2000review}.
These devices can also be used as polarization modulators by exploiting the Pockels effect in birefringent crystals, such as lithium niobate (LiNbO$_3$), where the application of an electric field induces a change in the refractive index of the material. 
Being birefringent, the ordinary and extraordinary refractive indices vary differently as a function of the applied voltage. 
Configurations using a single EOM placed directly in the optical path are called \textit{inline modulators}. 
Moreover, including EOMs in a Mach-Zehnder interferometer enables intensity modulation in a configuration commonly called the \textit{Mach-Zehnder modulator}.

Recently, these devices have been successfully used for quantum communication applications, where precise control of the quantum states of photons is required, namely their phase, polarization, and intensity. 
In particular, they have been extensively employed for Quantum Key Distribution (QKD) protocols, where this modulation is critical to encoding quantum information, enabling protocols such as the decoy-state BB84 \cite{bennett2014quantum, hwang2003quantum}.

A major drawback of both inline and Mach-Zehnder modulators is the bias drift resulting from temperature and environmental fluctuations, thus necessitating active stabilization of the temperature and bias voltage to maintain a stable operating point.
Moreover, inline polarization modulators require high operating voltages, coupled with the fact that the birefringence of the waveguide induces polarization mode dispersion (PMD), which reduces the degree of polarization for short pulses.

Sagnac loop-based modulators have recently been proposed as an effective alternative to conventional inline and Mach-Zehnder modulators. The common path geometry inherently compensates for reciprocal phase shifts, reducing sensitivity to environmental perturbations such as temperature fluctuations and mechanical vibrations.
The Sagnac interferometer configuration is exploited to achieve both intensity \cite{Roberts2018} and polarization modulation \cite{iPOGNAC, agnesi2019all} while avoiding the need for bias voltage control.

The Sagnac modulator consists of an optical loop (often realized by fiber optics), typically formed using a beam splitter (for intensity modulation) or a polarizing beam splitter (for polarization modulation) with the light entering the input port being split into two counter-propagating beams that travel the loop in opposite directions (see Fig. \ref{fig:sagnac-modulator}).
By placing an EOM asymmetrically within the loop, the co-propagating and counter-propagating beams arrive at the modulator at different times, allowing them to be modulated independently, thus controlling the relative phase between them.
However, this asymmetry also imposes a limit on the maximal repetition rate of optical pulse transmission. To prevent the counter-propagating component of the pulse $N$ from overlapping with the co-propagating component of the pulse $N+1$, it is necessary that the transmission frequency be lower than a value that depends on the propagation time of the pulse in the modulator and the asymmetry (i.e. its position in the Sagnac loop).
Overcoming this limitation by completely removing the asymmetry implies modulating both co-propagating and counter-propagating components at the same time, which presents a significant experimental challenge.

In this work, we introduce a complete model of the Sagnac modulator and propose some useful modulation techniques that can be used for both intensity and polarization modulation. 
Moreover, we present a study on the symmetric Sagnac configuration and how its challenges can be tackled with carefully chosen modulation schemes.

To provide a clear structure for this work, we briefly outline the organization of the paper below.
In Section \ref{sec:sagnac} we present a complete theoretical model for all Sagnac modulators, expanding further into the symmetric configuration, a non-standard configuration in the literature. 
In sections \ref{sec:polarization-sagnac} and \ref{sec:intensity-sagnac} we apply this theoretical model to two relevant applications, polarization and intensity modulation, respectively. 
Moreover, at the end of Section \ref{sec:intensity-sagnac} (Sec. \ref{sec:intensity-sagnac-symm}) we present a novel use of the symmetric Sagnac intensity modulator to carve optical pulses from a continuous wave laser.
In Section \ref{sec:RepRate} we present an analysis on the maximum achievable repetition rate in both the asymmetric and symmetric configurations, focusing on a limited but relevant number of modulation techniques for both.

\section{Principles of Sagnac modulators}
\label{sec:sagnac}

A traveling-wave electro-optic phase modulator is designed with an optical waveguide in an electro-optic material and an adjacent electrode structure forming a coplanar waveguide for a microwave signal.
This configuration allows the optical and electrical pulses to interact continuously along the length where the two waveguides are adjacent. 
To maximize this interaction, it is crucial to match not only their physical alignment but also their effective refractive indices (or phase velocities).
Proper velocity matching ensures phase synchrony between the optical and electrical waves throughout the device length, enabling effective modulation, as long as both the optical and electrical waves propagate in the same direction, i.e., co-propagating. 
This structure is particularly advantageous for high-speed applications, as it supports high repetition rates and multiple optical pulses can propagate while interacting primarily with their corresponding electrical pulse.
However, traveling-wave modulators can also operate when the optical pulses propagate in the opposite direction to the electrical signal.
In this counter-propagating configuration, the interaction between the two waves changes, resulting in a different modulation effect compared to the co-propagating configuration.
This effect arises from the relative motion between the optical and electrical pulses, which alters the effective interaction time and modulation efficiency.
Such a modulator imposes a phase $\phi_s(t)$ on the optical wave in response to an electrical signal $s(t)$, such that
\begin{equation}
\label{eq:phis2}
    \phi_s(t)=
    \int  e^{i\omega t}H(\omega)S(\omega)\dd\omega\,,
\end{equation}
where $H(w)$ is the transfer function of the modulator and 
\begin{equation}
    S(\omega)= \frac{1}{2\pi} \int e^{-i\omega t}s(t)\dd t\,
\end{equation} 
is the Fourier transform for the electrical signal $s(t)$.

Assuming that the velocities of the electrical and  optical waves are equal inside the phase modulator, the transfer function takes a simple but distinct form between the co-propagating (co) and counter-propagating (ct) cases, such that
\begin{equation}
\label{eq:Hmodulator}
\begin{aligned}
    H_{\rm co}(\omega)=\frac{\pi}{V_\pi}\,,\qquad
    H_{\rm ct}(\omega)=\frac{\pi}{V_\pi}{\rm sinc}(\tau_{\rm mod}\omega)\,,    
\end{aligned}
\end{equation}
where ${\rm sinc}(x)=\frac{\sin x}{x}$, $\taumod$ is the propagation time of the electrical (or optical) signal inside the modulator, and $V_\pi$ is the voltage of the electrical signal corresponding to a phase shift of $\pi$ \cite{alferness1982waveguide}. 
Typical values (which can vary depending on the specific commercial device used) are $V_\pi\simeq 5V$ and $\taumod\approx320$~ps. 
The previous relations are obtained for a simplified modulator model (neglecting electrical absorption and bandwidth saturation effects).
For simplicity in the modeling, we treat $V_\pi$ as a constant and neglect its frequency dependence (even though at high frequencies $V_\pi$ typically increases).

In the co-propagating configuration, the optical and electrical signals remain synchronized along the modulator, leading to a relatively uniform phase modulation. 
In contrast, in the counter-propagating case, the imposed phase depends on the angular frequency $\omega$ of the electrical signal, effectively acting as a low-pass filter. 
Because the two waves travel in opposite directions, their overlap occurs only for a very short period of time with respect to the co-propagating overlap time, during which the optical pulse accumulates the entire interaction with the electrical field. 
This short interaction window makes the counter-propagating scheme more sensitive to the physical length of the waveguide and can lead to less efficient modulation capabilities, implying a higher modulation voltage requirement to impose the same phase with respect to the co-propagating case.
We note that the asymmetry between co-propagating and counter-propagating case is more relevant when high-bandwidth electrical signal are considered: indeed, for frequencies $\omega$ much smaller that $1/\taumod$, the difference between the two transfer functions $H_{\rm co}(\omega)$ and $H_{\rm ct}(\omega)$ becomes negligible.

\begin{figure}[t!]
    \centering
    \includegraphics[width=\linewidth]{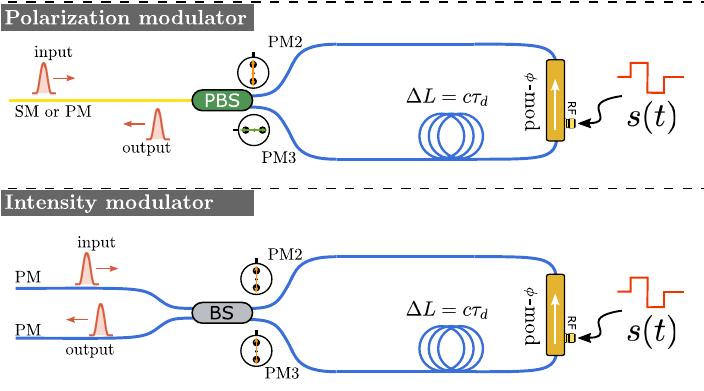}
    \caption{Polarization and intensity modulator based on the Sagnac scheme. $\phi$-mod indicates the electro-optic phase modulator, while the white arrow indicates the direction of the RF microwave signal. $\Delta L$ quantifies the asymmetry of the 
    phase modulator within the Sagnac loop. For symmetric configuration $\Delta L=0$.}
    \label{fig:sagnac-modulator}
\end{figure}

Combining the two transfer functions mentioned above, the model can be applied to describe the transfer function of schemes that use a modulator within a Sagnac loop, as illustrated in Fig. \ref{fig:sagnac-modulator}. 
For a Sagnac modulator, whether for intensity or polarization modulation, the resulting transfer function can be expressed by combining the two terms of Eq. \eqref{eq:Hmodulator} as
\begin{equation}
    \label{eq:transf-function}
    \Hsagnac(\omega) = \frac{\pi}{V_\pi} 
    e^{i\omega t_0}[1 - e^{-i\omega\taud} \text{sinc}(\taumod\omega)]\,,
\end{equation}
where $\taud = \Delta L/c$ is the time corresponding to the delay line $\Delta L$ and $t_0$ is the overall delay between the electrical and optical signal. 
Note that if the delay line $\Delta L$ is placed before or after the modulator, the sign of $\taud$ can be inverted. 

Through the transfer function $\Hsagnac(\omega)$, using \eqref{eq:phis2}, it is possible to calculate the phase as a function of time, $\phi_s(t)$, which will be applied to an optical signal when an electrical signal with a given temporal trend $s(t)$ is applied to the modulator. 
Alternatively, $\phi_s(t)$ can be evaluated by calculating the inverse Fourier transform of the transfer function (for simplicity, we neglect the overall delay $t_0$)

\begin{align}
\label{eq:hsagnac}
    \hsagnac(t) &= \int e^{i\omega t}\Hsagnac(\omega)d\omega \\ 
    &=\frac{2\pi^2}{V_\pi} \left[\delta_{\rm D}(t)
     - \frac{1}{2\taumod} \Pi \left(\frac{t-\taud}{2\taumod}\right)\right] \nonumber,
\end{align}
where $\delta_{\rm D}(t)$ is the Dirac-delta and $\Pi(t)$ is the rectangular function
\begin{equation}
    \Pi(t)=\begin{cases}
    1\,, & |t|\leq 1/2
    \\
    0\,, & |t|>1/2\,.
    \end{cases}
\end{equation}

The phase $\phi_s(t)$
 is therefore obtained from the convolution
\begin{equation}
\label{eq:phis_by_convolution}
\begin{aligned}
    \phi_s(t) &= \frac{1}{2\pi} \int s(t')
    \hsagnac(t - t')dt'
    \\&=\phi_{co}(t)-\phi_{ct}(t)\,,
\end{aligned}
\end{equation}
where
\begin{equation}
\begin{aligned}
\phi_{co}(t)&=\frac{\pi}{V_\pi}s(t)\,,
\\
\phi_{ct}(t)&=
\frac{\pi}{V_\pi}\frac{1}{2\taumod} \int_{t-\taumod}^{t+\taumod} s(t'-\taud)dt'
\end{aligned}
\end{equation}
represent, respectively, the phase applied to the co-propagating and counter-propagating components, where the latter behaves effectively as the moving average of the electrical signal around $\taud$.
The two expressions \eqref{eq:phis2} and \eqref{eq:phis_by_convolution} are equivalent and can be used to calculate $\phi_s(t)$  starting from the shape of the electrical signal 
$s(t)$. 

\subsection{Rectangular driving signal}
The asymmetry between co-propagating and counter-propagating optical signals can be effectively illustrated in the behavior of the modulator when a rectangular wave signal (the typical signal used in the modulators described previously) of duration $\taus$ is given as the input. 
For a rectangular signal with maximum voltage $V_0$
\begin{equation}
\label{eq:square-pulse}
    s(t)=V_0
    \Pi\left(\frac{t-\taus/2}{\taus}\right)\,,
\end{equation}
the expression for $\phi_{ct}(t)$ in \eqref{eq:phis_by_convolution} can be simplified to a trapezoidal equation
\begin{equation}
    \phi_{ct}(t)=
    \pi\frac{V_0}{V_\pi}\cdot\begin{cases}
    0 & t'\leq -\taumod
    \\
    \min(\frac{t'+\taumod}{2\taumod},\frac{\taus}{2\taumod}) & -\taumod\leq t' \leq \taumod
    \\
    \min(1,\frac{\taumod+\taus-t'}{2\taumod}) & 
    \taumod\leq t' \leq \taumod+\taus
    \\
    0 & t'\geq \taumod+\taus
    \end{cases}
\end{equation}
with $t'=t-\taud$.

\begin{figure}[t!]
    \centering
    \includegraphics[width=\linewidth]{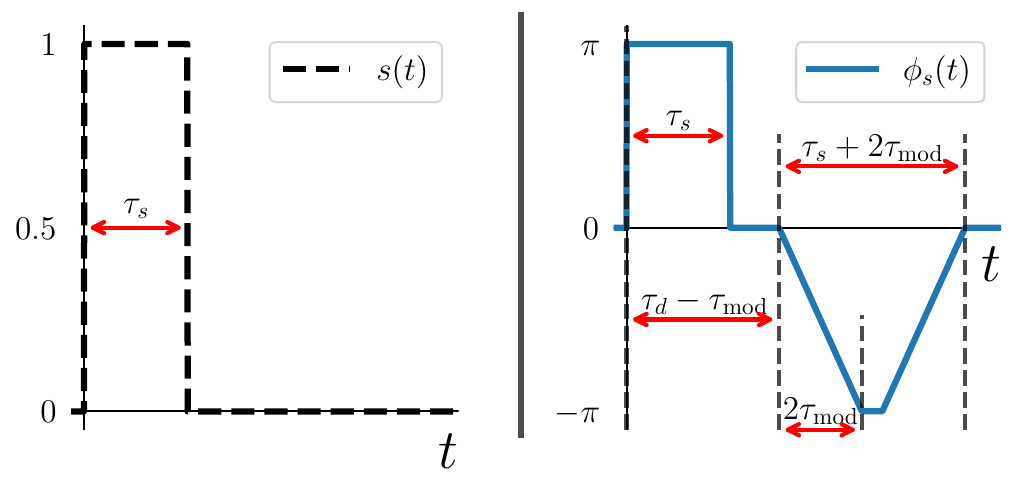}
    \caption{Phase $\phi_s(t)$ for rectangular driving signal $s(t)$. We considered the case $\taud\geq \taus+\taumod$ and $\taus\geq2\taumod$.
    The signal $s(t)$ is reported in units of $V_0$, while $\phi_s(t)$ in given in unit of $\frac{V_0}{V_\pi}$.
    For simplicity we set $t_0=0$.
}
    \label{fig:phis_asymm}
\end{figure}

To provide a clearer understanding of the output phase $\phi_s(t)$ in response to a rectangular driving signal $s(t)$, Fig. \ref{fig:phis_asymm} shows a representative example corresponding to cases $\taud -\taumod\geq \taus  $ and $\taus \geq 2\taumod$. 
It can be noted that the phase applied to the co-propagating signal exactly follows the shape of the rectangular wave input signal. 
For the counter-propagating signal, the phase is modified by the transfer function, resulting in a trapezoidal shape. 

If $\taus>2\taumod$, the phase $\phi_{ct}$ increases linearly from the instant $t=t^*\equiv\taud-\taumod$,  reaching its maximum value $\pi V_0/V_\pi$ at time $t=t^*+2\taumod$. 
From this instant on, the phase remains constant for a time $\taus-2\taumod$ up to the point $t  =t^*+\taus $ from which it decreases linearly to 0 at time $t =t^*+\taus + 2\taumod$.

If instead the duration of the electrical signal $\taus$ satisfies $\taus\leq 2\taumod$, the phase $\phi_{ct}$ increases to $\pi\frac{V_0}{V_\pi}\frac{\taus}{2\taumod}$ at time $t=t^*+\taus$, remains constant until $t=t^*+2\taumod$ when it decreases to $0$ at $t=t^*+\taus+2\taumod$. 
In general, the maximum phase applied to the counter-propagating optical signal can be written as
\begin{equation}
\label{eq:phicrmax}
    \phi^{\rm max}_{ct}=\frac{V_0}{V_\pi} \pi \cdot \min\left(1, \frac{\taus}{2\taumod}\right)\,.
\end{equation}

It is important to emphasize that to achieve the maximum phase modulation under the condition of $\taus \ge 2 \taumod$, the counter-propagating electrical pulse must transfer enough energy to the optical signal along the entire length of the modulator, similar to what occurs in the co-propagating configuration. 
To fulfill this condition, the optical signal must enter the waveguide after the electrical signal; the latter must have propagated through the entire modulator length, which requires a time \(\tau_{\text{mod}}\). 
Moreover, the electrical signal must persist for an additional duration of \(\taumod\), corresponding to the transit time of the optical wave through the modulator. 
Consequently, the minimum total duration over which the electrical signal must be active is \(2\tau_{\text{mod}}\). 
This can also be considered as the minimum condition that allows to inscribe the co-propagation rectangular signal into a counter-propagation trapezoid with the same height. 
When injecting an optical pulse of width $\tauop$ into the Sagnac modulator, to impose a constant phase over the entire pulse, it must hold that $\tauop\le\taus$ if the optical pulse is synchronized with  $\phi_{co}(t)$, while $\tauop\le\taus-2\taumod$ if the optical pulse is synchronized with $\phi_{ct}(t)$. In this case, we can reduce the asymmetry $\taud$ to the lower bound $\taud\geq\tauop+\taumod$.

It is worth noting that the particular example of a rectangular electrical signal is an idealization and does not fully represent a real, imperfect, implementation.
However, the model as presented can take some of these imperfections into account.
For example, any electrical waveform generator is limited in bandwidth, which can be easily taken into account by modifying the signal $s(t)$ according to said limitation. The same condition can apply, for example, to bandwidth limits on electrical amplifiers.
Another common limitation, caused by imperfect matching of the indices of refraction of both the electrical and optical waveguide of the EOM, is the bandwidth limit within the EOM itself. This effect limits the modulator bandwidth typically up to $10$ to $20$ GHz.
Taking this effect into account implies modifying the transfer function presented in Eq. \eqref{eq:transf-function} appropriately.

\subsection{Symmetric configuration}\label{sec:sagnac-symm}

An interesting case arises when the Sagnac interferometer is operated in a symmetric configuration (i.e., $\taud=0$). 
In this regime, the asymmetry between co-propagating and counter-propagating interactions discussed in the previous section can be exploited to impose different phase shifts on the two optical components. 
While both directions are modulated by the same electrical signal, their responses do not cancel out as the counter-propagating component travels against the electrical pulse, thus experiencing a different effective modulation voltage than the co-propagating component.

In the symmetric configuration, the co- and counter-propagating signals are allowed to interfere. Due to the modulator’s asymmetry, the amplitude of the counter-propagating component also depends on the duration of the electrical pulse. Therefore, it is important that the electrical pulse satisfies $\taus < 2\taumod$ to prevent the two effects from canceling out within the modulation region. Furthermore, since in this scheme the electrical pulse can be made much shorter, it is advantageous to optimize throughput by choosing $\taus \approx \tauop$ when performing optical pulse modulation.

If a standard rectangular electrical pulse, such as the one defined in Eq.~\eqref{eq:square-pulse}, is used under the condition $\taus < 2\taumod$, the co- and counter-propagating contributions interfere destructively. 
In this case, since $\phi^{\rm max}_{co}=\pi\frac{V_0}{V_\pi}$ and $\phi^{\rm max}_{ct}=\phi^{\rm max}_{co} \frac{\taus}{2\taumod}$ (see eq. \eqref{eq:phicrmax}) the imbalance between the two effects can be pre-compensated by increasing the modulation voltage $V_0$ by a gain factor of
\begin{equation}
\label{eq:gain-factor}
    g_V=\left(1 - \frac{\phi_{ct}^{\rm max}}{\phi_{co}^{\rm max}}\right)^{-1}= \frac{2 \taumod}{2 \taumod - \taus} .
\end{equation}
Assuming a sensitive value for commercially available EOMs of $\taumod=320$~ps and the minimum achievable $\taus$ using a 6 GSa/s digital-to-analog converter ($\taus=166.6$~ps), this corresponds to a gain factor of $g_V\approx1.35$.
This increase in amplification may require cascades of amplifiers that also introduce distortion, decreasing the signal-to-noise ratio.

In this work, we propose a new option for the electrical signal $s(t)$ that overcomes the need for an increase in voltage and fully exploits the integration-like behavior of the counter-propagating component.
Thus, this new modulation scheme, defined as \textit{balanced modulation}, uses electrical signals defined as
\begin{equation}
    \label{eq:balanced-mod}
    s(t)=V_0\left[\Pi\left(\frac{t-\taus/2}{\taus}\right)-\Pi\left(\frac{t-3\taus/2}{\taus}\right)\right].eve
\end{equation}
This method requires the use of three voltage levels, implying an extra requirement of a digital-to-analog converter or equivalent circuit capable of generating all levels (e.g., two separate voltage signals, an RF voltage inverter coupled with an RF power combiner).
This approach can be used to improve state of the art schemes based on asymmetric Sagnac loops such as, for example, the intensity modulator proposed in \cite{Roberts2018}, the time-bin encoders proposed in \cite{scalcon2025low} and \cite{vijayadharan2025sagnac}, or the polarization encoder proposed in \cite{iPOGNAC}.

Moreover, the proposed modulation scheme is applicable to any architecture that exploits a double pass (in opposite directions) of light in the modulator, such as the scheme proposed in \cite{lucio2009proof}.

\section{Sagnac polarization modulators}\label{sec:polarization-sagnac}

\begin{figure*}[t]
    \centering
    \includegraphics[width=0.98\linewidth]{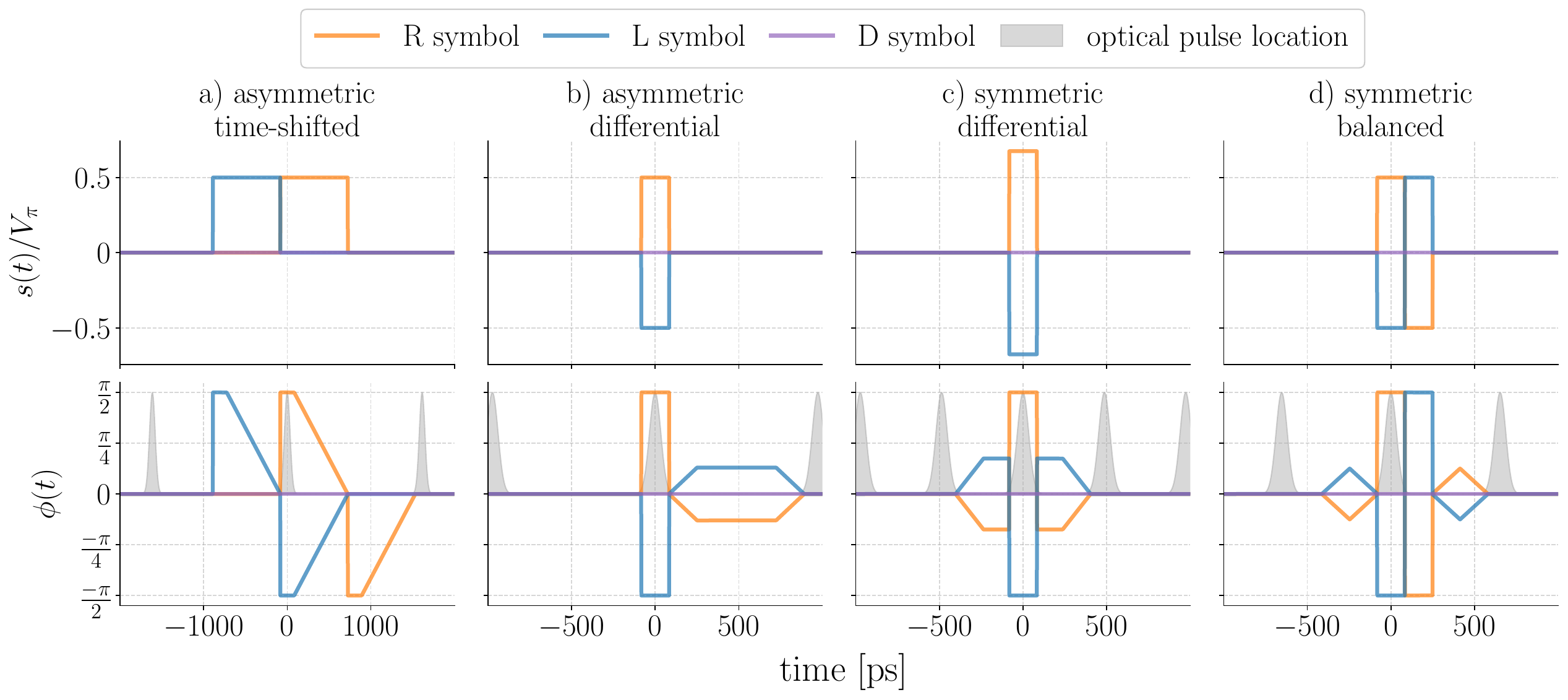}
    \caption{Modulations proposed in this work, characterized by signal $s(t)$ (top) and phase response $\phi(t)$ of the configuration (bottom). From left to right: a) time-shifted modulations for the asymmetric configuration (for    $\taus=2\taumod+\tauop$);
    b) differential asymmetric modulations (for $\taus=\tauop$);
    c) differential symmetric modulation (for $\taus=\tauop$);
    d) balanced symmetric modulations (for $\taus=\tauop$).
    For the asymmetric configuration we choose $\taud=\tauop+\taumod$, while for the symmetric configuration we have $\taud=0$.
    It should be noted that the voltage required to apply a $\pi/2$ phase on the optical pulse for the symmetric differential configuration is higher than all others by a gain factor of $g_V$ defined in Eq. \eqref{eq:gain-factor}. The locations of the pulses have been chosen to maximize the repetition rate.}
    \label{fig:modulations}
\end{figure*}

Any Sagnac polarization modulator \cite{agnesi2019all, iPOGNAC, wang202510, Li2020} is based on a Sagnac interferometer with an EOM inside the loop, where the standard beamsplitter is replaced by a polarizing beam splitter (PBS) to separate the co- and counter-propagating components according to the input polarization state (Fig. \ref{fig:sagnac-modulator}a). 
For a given electrical signal $s(t)$, an optical pulse is prepared in a diagonal polarization state $\ket{D}=\left(\ket{H} + \ket{V}\right)/\sqrt{2}$ (or any balanced superposition of $\ket H$ and $\ket V$). When this pulse is injected into a Sagnac modulator, the resulting output state is described by

\begin{equation}
\label{eq:ipognac-output-state}
\ket{\psi_{\rm out}(t)}=\frac{1}{\sqrt{2}}(|H\rangle + e^{i\phi_s(t)}|V\rangle)\,,
\end{equation}
which corresponds to any state belonging to the $X-Y$ equatorial plane of the Bloch sphere. 
This section focuses on a specific implementation of iPOGNAC \cite{iPOGNAC} tailored for QKD applications where only three polarization states are required to realize the three-state BB84 protocol \cite{PhysRevA.98.052336}.

In previous works, Sagnac polarization modulators have been used with a carefully chosen asymmetry in the Sagnac interferometer, allowing for what we will refer to as \textit{time-shifted modulation}.
As described in Section~\ref{sec:sagnac}, when using an asymmetric iPOGNAC with a delay line $\taud \ge \taumod + \taus$, asymmetric modulation is achieved by sending a sufficiently large electrical pulse ($\taus \ge 2\taumod$) to modulate with the same intensity the co-propagating and counter-propagating optical pulses of width $\tauop\le\taus-2\taumod$. 
The three symbols are achieved by sending an electrical pulse aligned with the optical co-propagating for $\ket{L}$, shifting the electrical symbol by $\taud$ for $\ket{R}$, and sending no electrical pulse for $\ket{D}$. Thus, we can define the corresponding $s_i(t)$ required to generate each symbol, with $i=R,L,D$ as
\begin{equation}
    \begin{split}
        s_L(t)&=\halfwavev\Pi \left(\frac{t-t_0}{\taus}\right),\\
        s_R(t)&=\halfwavev\Pi \left(\frac{t-t_0-\taud}{\taus}\right),\\
        s_D(t)&= 0,
    \end{split}
\end{equation}
where $\halfwavev$ is the half-wave voltage of the phase modulator and $t_0$ is an arbitrary time such that the phase electrical signal aligns with the optical pulse (Fig. \ref{fig:modulations}a).

The asymmetric configuration coupled with the time-shifted modulation allows the creation of three polarization states by only employing two-level signals (i.e. $V=\{0, \halfwavev\}$).
This configuration, however, implies a limitation on the maximum achievable repetition rate $R$ of the modulation (see Section \ref{sec:RepRate}). 

If more than two voltage levels are available (in particular, $\halfwavev$, 0, and $-\halfwavev$), it is possible to increase the repetition rate by having the electrical pulse aligned with the co-propagating optical pulse in both cases, which we will refer to as \textit{differential modulation} from now on, such that 
\begin{equation}
    \begin{split}
        s_L(t)&=\halfwavev\Pi \left(\frac{t-t_0}{\taus}\right),\\
        s_R(t)&=-\halfwavev\Pi \left(\frac{t-t_0}{\taus}\right),\\
        s_D(t)&= 0,
    \end{split}
\end{equation}
which also removes the requirement for $\taus\ge2\taumod$ (Fig. \ref{fig:modulations}b), reducing it instead to $\taus\ge\tauop$ so that the entire optical pulse is modulated equally.
The same modulation scheme can also be used in a symmetric configuration as presented in Section \ref{sec:sagnac-symm}, requiring an increase in the modulation voltage by a gain factor as described by Eq. \eqref{eq:gain-factor} (Fig. \ref{fig:modulations}c).

However, it is possible to remove the requirement for a voltage increase by using \textit{balanced modulation} (Fig. \ref{fig:modulations}d), which has electric signals with a shape given by Eq. \eqref{eq:balanced-mod}, such that
\begin{equation}
    \begin{split}
        s_L(t)&=\halfwavev\left[\Pi\left(\frac{t-t_0}{\taus}\right)-\Pi\left(\frac{t-t_0-\taus}{\taus}\right)\right],\\
        s_R(t)&=-\halfwavev\left[\Pi\left(\frac{t-t_0}{\taus}\right)-\Pi\left(\frac{t-t_0-\taus}{\taus}\right)\right],\\
        s_D(t)&= 0.
    \end{split}
\end{equation}
This modulation scheme not only relaxes the requirement for voltage amplification that is otherwise required, but it also has the property of each symbol being of zero-average voltage. 
In standard commercial AC-coupled amplifiers, the amplification factor is usually dependent on the average signal level (i.e. the DC component).
By each symbol being zero-average voltage, the average signal level is kept at zero regardless of the sequence being sent, greatly mitigating patterning effects.

\subsection{Experimental validation}

\begin{figure}
    \centering
    \includegraphics[width=1\linewidth]{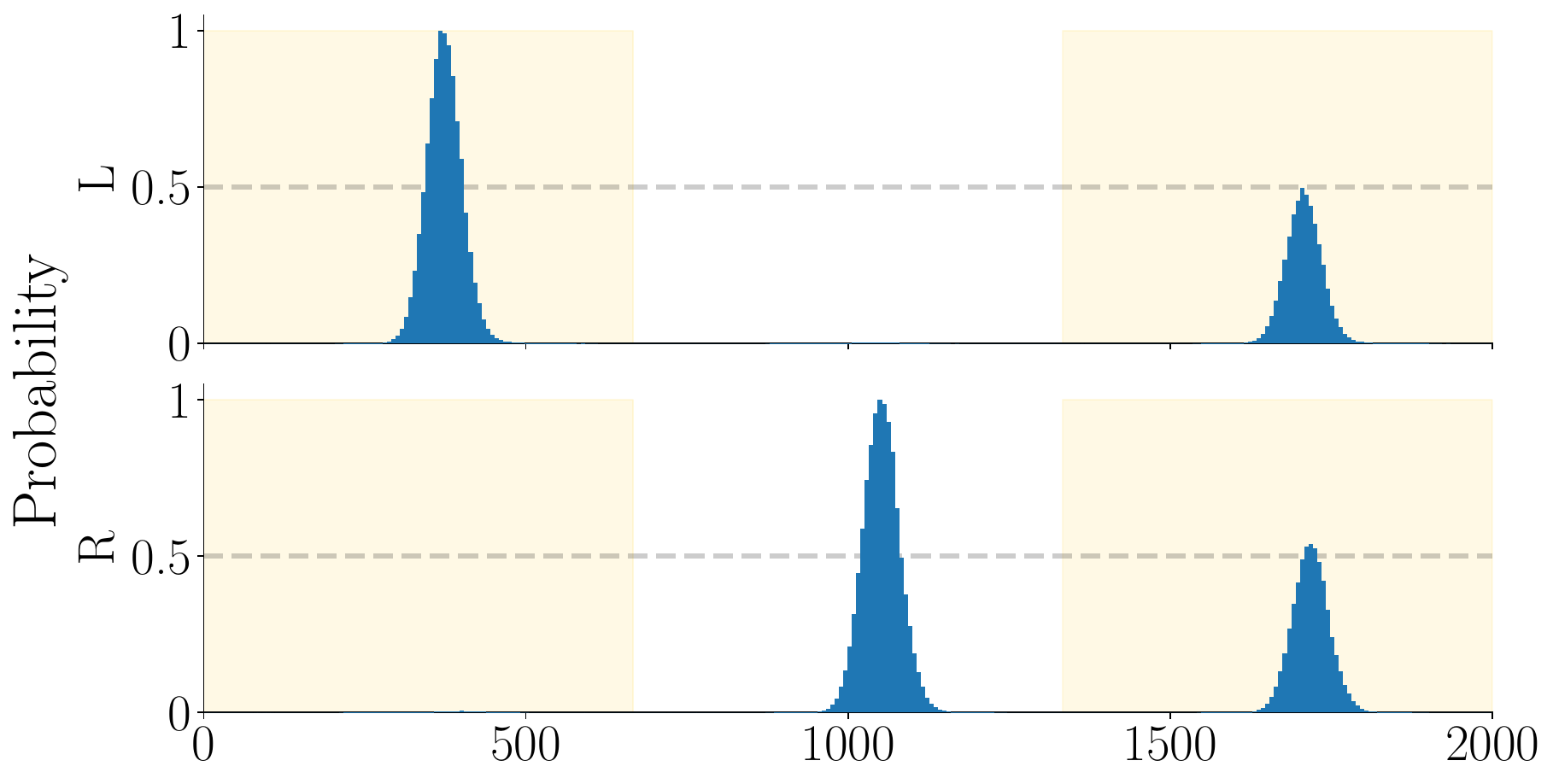}
    \caption{Generated states at $1.5$~GHz of repetition rate using the symmetric balanced configuration and projected on the $Y$ basis, with a polarization extinction ratio of $\rm{PER}_Y=23.147\pm0.003$~dB observed between the $\ket{L}$ and $\ket{R}$ states.
    }
    \label{fig:1.5GHz}
\end{figure}

The balanced modulation scheme was tested on a Sagnac polarization modulator by injecting pulsed $\ket{D}$ states at a repetition rate of $R=1.5$~GHz. By creating all the required states for the three-state BB84 protocol ($\ket{L}$, $\ket{R}$ and $\ket{D}$), and projecting into the $Y$ and $X$ bases, a polarization extinction ratio ($\rm PER$) for each basis was obtained as $\rm{PER}_Y=23.147\pm0.003$~dB and $\rm{PER}_X=23.981\pm0.005$~dB  (Fig. \ref{fig:1.5GHz}). 
If this state preparation was used for QKD applications, for example, the intrinsic $\text{QBER}$ can be obtained as 
\begin{equation}
    \rm{QBER}=\frac{1}{1+10^{\rm{PER}/10}},
\end{equation}
which would correspond to $\rm{QBER}_Y\approx0.5$\% and $\rm{QBER}_X\approx0.4$\%. 

\begin{figure}[t]
    \centering
    \includegraphics[width=1\linewidth]{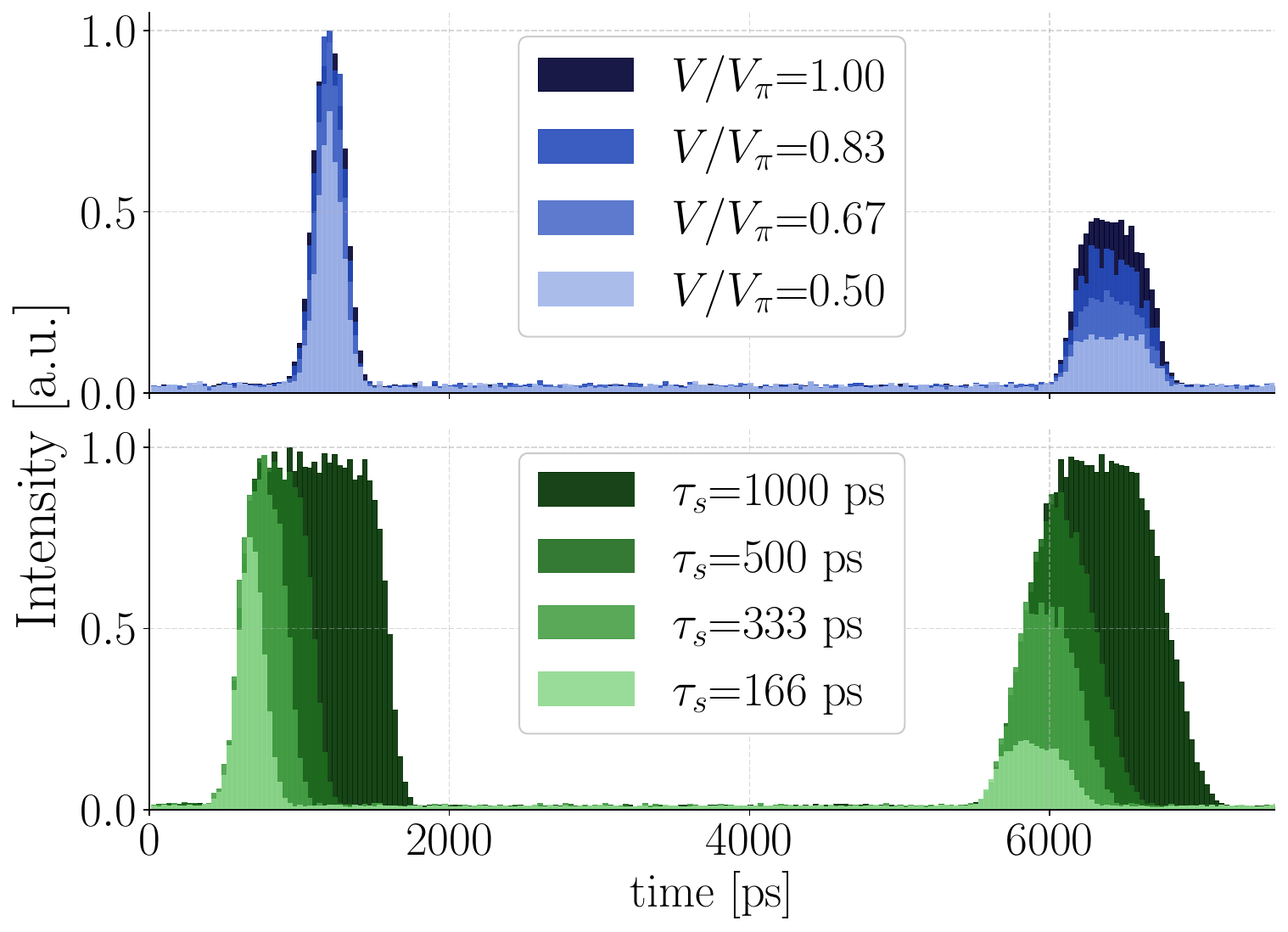}
    \caption{Measured intensity at the output of an asymmetric ($\taud=5$~ns) iPOGNAC when injecting continuous wave light in a fixed $\ket{D}$ polarization state and projecting in the $\ket{A}$ state, driven by a rectangular electrical signal when varying $\tau_s$ (bottom) and $V$ (top).}
    \label{fig:std-asymm-vs}
\end{figure}

In first instance, to validate the phase response predicted by Eq. \eqref{eq:phis_by_convolution}, an asymmetric ($\taud=5$~ns) iPOGNAC system was used, driven by a rectangular electrical signal, while injecting continuous wave laser light with a defined $\ket{D}$ polarization to its input port.
Then, the projection of the output state on the $\ket{A}$ polarization was measured.
Assuming an output state like Eq. \eqref{eq:ipognac-output-state}, the probability of measuring the system in the state $\ket{A}$ can be written as 
\begin{equation}
\begin{split}
    P_A(t) &= \left| \braket{A|\left(\ket{H} + e^{i\phi_s(t)}\ket{V}\right)} \right|^2 \\
    &= \sin^2\left(\frac{\phi_s(t)}{2}\right).
\end{split}
\end{equation}
This probability corresponds to the projection onto the chosen state, which can be generalized to any arbitrary measurement basis. 
In the case of a continuous optical signal, this projection translates into an intensity modulation, effectively carving the light field according to the applied modulation. 
Figure~\ref{fig:std-asymm-vs} shows the relative intensity measured at the output (proportional to $P_A(t)$) as a function of $\tau_s$ and the modulating voltage, for an asymmetry of $\tau_d = 5$~ns.
We note that in Fig.~\ref{fig:phis_asymm}(right) we reported $\phi_s(t)$ (which could be negative), while in Fig.~\ref{fig:std-asymm-vs} we showed an intensity proportional to $\sin^2(\phi_s(t)/2)$ (always positive).

Finally, to compare both modulation schemes applicable to symmetric Sagnac modulators, an iPOGNAC system was used with both differential and balanced modulation, in the same conditions previously mentioned.
To take into account the imperfections of the system, the electrical signals $s^*(t)$ were measured using an RTP164B oscilloscope by Rohde \& Schwarz after the corresponding amplification stage. 
The maximum voltage used for both cases remained constant to highlight the reduction in applied phase, discussed in Section \ref{sec:sagnac-symm}. 
From the measured electrical signals $s^*(t)$, the phase response $\phi^*_s(t)$ is obtained following Eq. \eqref{eq:phis_by_convolution}, and the predicted probability of measuring the system in the state $\ket{A}$ can be obtained as $P_A(t)=\sin^2\left(\phi^*_s(t)/2\right)$.
Fig.~\ref{fig:std-asymm-vs}a shows the measured relative intensity at the output port of the iPOGNAC using Superconductive Nanowire Single Photon Detectors (SNSPD) (blue bars) compared to the one predicted by the model from the measured electrical signals $s^*(t)$. 
In first instance, it was assumed the predicted intensity was detected with ideal detectors, meaning no detection jitter (dashed purple line), which presented some discrepancies with respect to the measured intensity. 
Then, a Gaussian filter was used on the predicted intensity to simulate the behavior of the detection jitter of the SNSPDs (solid purple line). We note that the measured behaviour follows the prediction of the model reported in Fig.~\ref{fig:phis_asymm} when a finite bandwidth of the electrical signal and SNSPD jitter are considered.

\begin{figure*}[t]
    \centering
    \includegraphics[width=0.9\linewidth, keepaspectratio]{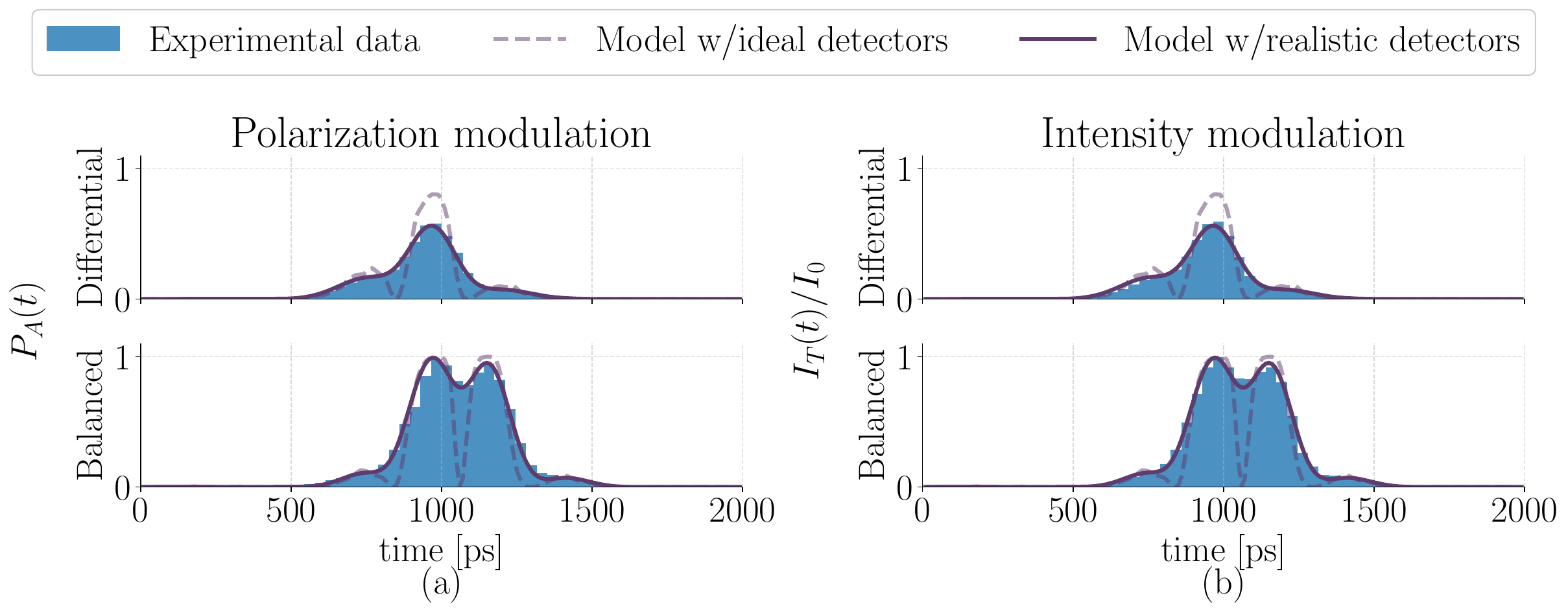}
    \caption{Comparison between the differential and balanced symmetric modulation for (a) polarization modulator and (b) intensity modulator. In dashed purple, the expected measuring probability assuming a non-ideal electrical signal (directly measured) and ideal detectors (no jitter). In solid purple, the expected measuring probability now taking into account the SNSPD jitter.}
    \label{fig:bal-vs-diff}
\end{figure*}

\section{Sagnac intensity modulators}\label{sec:intensity-sagnac}
Sagnac intensity modulators consist of a Sagnac interferometer with an EOM inside the Sagnac loop (Fig. \ref{fig:sagnac-modulator}b). When used in the pulsed regime, as presented by Roberts et al. \cite{Roberts2018}, this scheme allows for controllable attenuation of  optical pulses by carefully choosing the modulating voltage and the beamsplitter ratio. 

Assuming a time-independent beamsplitter with transmittance $T$, for a given time $t$, the output intensity of a Sagnac interferometer will be given by 
\begin{equation}
    I_T(T,t)= I_0\left| e^{i \phi(t)} (T - 1) + T\right|^2,
\end{equation}
with  $I_0$ the intensity at the input port and $\phi(t)$ the phase difference between the two arms of the Sagnac at time $t$. Combining this with Eq. \eqref{eq:phis_by_convolution}, the time-dependent transmission can be obtained, from which the extinction ratio $ER$ can be estimated. Assuming $\phi_s(t)=0$ for a given time $t$, and the absolute maximum phase to be $\phi_{\rm max}=\max|\phi_s(t)|$, the extinction ratio is given by
\begin{equation}
    ER(T) = 10 \log_{10} \left| \frac{e^{i \phi_{\rm max}} (T - 1) + T}{2T-1}\right|^2.
\end{equation}

As highlighted by Roberts et al., to reduce patterning effects on the system, it proves convenient to set $\phi_{\rm max}=\pi$, such that the modulation occurs at the peak of the transfer function, where the derivative is zero, reducing small fluctuations of the electrical voltage that could occur in the input electrical chain. It is worth noting that to properly attenuate an optical pulse, the phase $\phi_{\rm max}$ has to be maintained for the entire duration of the pulse.

Once again, to provide further experimental validation of the proposed model, the same procedure presented in the previous section was performed using a Sagnac intensity modulator with a beamsplitter with transmittance $T=1/2$, maintaining the same electrical signals $s^*(t)$. 
For this measurement, instead of projecting to a particular polarization state, we simply measured the output intensity of the modulator, such that
\begin{equation}
\label{eq:transmittance-sagnac}
\begin{split}
        I_T(T, t)& =I_0\left| e^{i \phi^*_s(t)} (T - 1) + T\right|^2\\
        &=I_0\sin^2\left(\frac{\phi^*_s(t)}{2}\right),
\end{split}
\end{equation}
showing the same time dependence as for the polarization case (Fig. \ref{fig:bal-vs-diff}b), with, as before, $I_0$ the intensity at the input port.

\subsection{Pulse generation with symmetric configuration}\label{sec:intensity-sagnac-symm}

\begin{figure}
    \centering
    \includegraphics[width=1\linewidth]{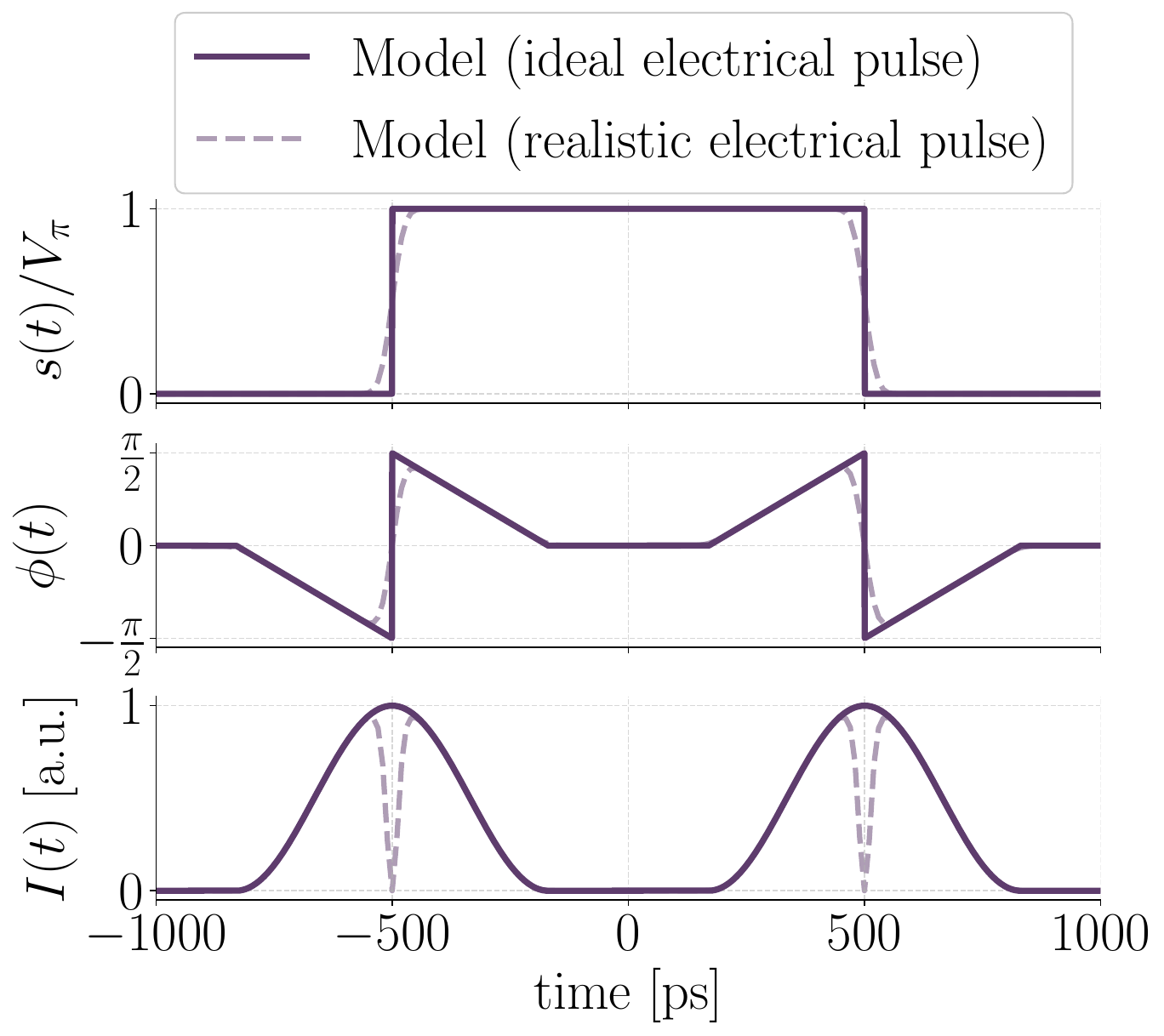}
    \caption{Example of pulse generation through carving using the Sagnac intensity modulator in symmetric configuration. Despite the realistic scenario presenting a ``dip" in the center of the pulse, it is only when $\taud<\taumod$ that the intensity reaches zero, otherwise resulting similarly to the ideal case.}
    \label{fig:pulsegenerationsymm}
\end{figure}

When using a Sagnac intensity modulator in the symmetric configuration, everything exposed in section \ref{sec:sagnac-symm} is still valid, allowing for an increase in the repetition rate of the intensity modulation when using pulsed light at the input. 
Moreover, in addition to this, a new application arises from the nature of Sagnac interferometers: \textit{pulse generation} by carving from continuous wave light (demonstrated in \cite{vijayadharan2025sagnac}). 
For this regime to occur, the system must be used in the condition $\taus>\taud$, which occurs automatically in the symmetric configuration, where $\taud=0$.

Assuming an input signal of the form $s(t)=\Pi\left(t/\taus\right)$, it can be seen that the response function of the phase modulator, given by eq. \eqref{eq:phis_by_convolution}, is nonzero only at the rising and falling edge of the electrical pulse (Fig. \ref{fig:pulsegenerationsymm}). 
When reaching the beamsplitter at the output, only those nonzero phase points will interfere constructively on the second mode of the beamsplitter, thus generating two optical pulses separated by $\taus$. 
However, when taking into account imperfections in the electrical signal (i.e. not perfectly square), the transition from negative to positive phase difference is `smoothed', thus creating a `dip' on the optical pulse due to the transition having a zero-phase-difference point on each of the edges.
If pulse shaping is relevant, this can be easily solved by adding a non-zero delay of at least $\taumod$, such that the zero-phase-difference points are removed. 

\begin{figure}
    \centering
    \includegraphics[width=1\linewidth]{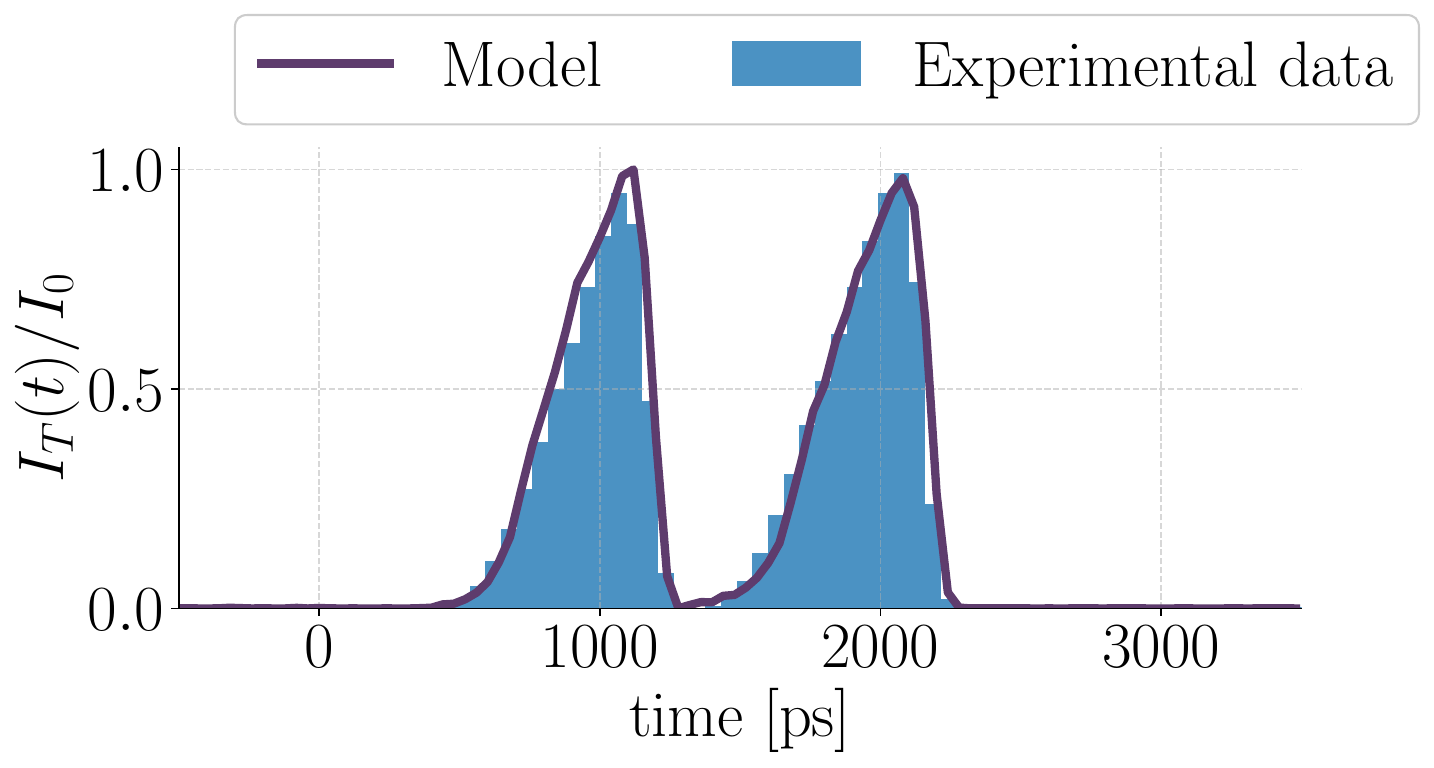}
    \caption{Generated optical pulses using a slightly-asymmetric Sagnac modulator with $\taud\approx\taumod$ and $\taus=1$~ns. In purple, the expected measuring probability assuming a non-ideal electrical signal.}
    \label{fig:pulsegenerationdata}
\end{figure}

To validate the proposed scheme, we implemented a slightly-asymmetric Sagnac modulator ($\taud\approx\taumod$) to carve two optical pulses separated by $\taus=1$~ns.
As described in Section \ref{sec:polarization-sagnac}, the electrical signals $s^*(t)$ were measured using an oscilloscope after the corresponding amplification stage, from which the phase response $\phi_s^*(t)$ was obtained according to Eq. \eqref{eq:phis_by_convolution}.
We then compared the generated optical pulses with the expected transmittance of the modulator as described in Eq. \eqref{eq:transmittance-sagnac}, highlighting once again the validity of the model (Fig. \ref{fig:pulsegenerationdata}).

\section{Exploring the high repetition rate limits}
\label{sec:RepRate}
Selecting an appropriate modulation scheme is crucial for maximizing the source repetition rate. Modulation strongly depends on the geometry of the modulated system: for example, a loop scheme (like the one of the iPOGNAC) has different optimized modulations than an inline system. 
This is because, as shown in the previous section, additional inter-symbol interference needs to be taken into account.
Here, three different modulation schemes will be covered, one related to an asymmetric Sagnac configuration (time-shifted modulation, see Sec. \ref{sec:polarization-sagnac}), one to a symmetric one (balanced modulation, see Sec. \ref{sec:sagnac-symm}), and one valid for both (differential modulation, see Sec. \ref{sec:polarization-sagnac} and \ref{sec:sagnac-symm}). 

For each modulation format, performance is characterized by two metrics: the symbol period $\tsym$, which is defined as the minimum time required to concatenate two symbols without experiencing inter-symbol interference in the modulation region, and the normalized voltage range $\Delta V = (V_{\text{max}} - V_{\text{min}})/V_{\pi}$  required to achieve all the symbols.
For the sake of absolute maximum ratings, we will assume that the target optical pulse is smaller than $\taus$.
These parameters are summarized in Table \ref{tab:max-rates}, where the modulator length was experimentally estimated as $\taumod=320$~ps and the minimum achievable electrical signal width as $\taus=166.6$~ps to provide a numeric value for all cases. 
\begin{table*}[t]
    \centering
  \begin{tblr}{colspec={l c c c c}, row{1}={font=\bfseries}, row{even}={bg=gray!10}}
        Modulation scheme& $\Delta V$ & $\tsym$& $\taus^*$~[ps]  & $t^*_{sym}$~[ps] & $R_{\rm max}~[GHz]$\\
    \toprule
         Time-shifted            & 0.5    & $2(2\taumod+\tauop)$   & $640.0$ & $1613.3$ & $\sim0.62$  \\
         Differential asymmetric & 0.5    &  $2(\taumod+\taus)$ & $166.6$ & $973.2$   & $\sim1.03$  \\ 
         Differential symmetric  & 0.685  &  $\taumod+\taus$    & $166.6$ & $486.6$   & $\sim2.06$  \\
         Balanced                & 0.5    &  $\taumod+\taus$   & $2\cdot166.6$ & $653.2$   & $\sim1.53$  \\
    \bottomrule
  \end{tblr}
  \caption{Performance parameters for all implemented modulations, with $\Delta V$ the normalized range of voltage required to achieve all states for the three-state BB84 protocol and $\tsym$ the minimum period as a function of the setup parameters.
  The minimum period $\tsym^*$ and maximum repetition rate $R_{\rm max}$ were obtained assuming the minimum electrical signal duration $\taus^*$ for a optical pulse duration of $\tauop=166.6$~ps and a modulator length of $\taumod=320$~ps.
  }
  \label{tab:max-rates}
\end{table*}

With each of the proposed schemes, it is possible to increase the repetition rate of the system at the expense of non-ideal modulation, that is, allowing inter-symbol interference. 
In this regime, the total time required for the response function to a given $s(t)$ is greater than the repetition rate desired for the modulation. 
Depending on the desired use of the system, this behavior can be tolerated to allow an increase in the overall repetition rate. 
When using periodic modulation signals, this behavior can be pre-compensated, taking into account the inter-symbol interference terms. 
However, for applications like QKD where the modulating signal has to be randomly chosen at the qubit repetition rate, this interference leads to randomly distributed imperfect modulation, resulting in a net increase in QBER.

\subsection{Time-shifted modulation}
Time-shifted modulation is achieved when selectively targeting either the co-propagating or counter-propagating optical pulse in an asymmetric Sagnac configuration as described in Sec. \ref{sec:polarization-sagnac}. 
To avoid consecutive symbols from interfering, it is enough, from a conservative standpoint, to avoid overlap in the phase response.
As shown in Fig.~\ref{fig:phis_asymm}, for a given electrical pulse of width $\taus$, the total duration of the modulation is determined by both the co-propagating and counter-propagating phase responses. In general, assuming the asymmetry condition $\taud \geq 2\taumod$, the total modulation time for a symbol is $\tsym = \taus + \taud + \taumod$, where $\taud$ is the delay between the co- and counter-propagating contributions, and $\taumod$ is the modulator length.

To avoid overlap between the co-propagating phase response $\phi_{co}$ and the counter-propagating response $\phi_{ct}$, the delay must satisfy $\taud \geq \taus + \taumod$, which ensures that the counter-propagating phase starts only after the co-propagating phase ends. Substituting this condition into the expression for $\tsym$ gives the minimum period $\tsym \geq 2(\taus + \taumod)$. A convenient boundary condition is to set $\taus \geq 2 \taumod$, which ensures a sufficiently long pulse relative to the modulator transition time. In this case, the minimum separation between consecutive optical pulses is $\tsym \geq 6 \taumod$.
However, if the finite duration of the optical pulses of length $\tauop$ is considered, it is possible to further optimize the repetition rate. 
This is because the electrical modulation window is strictly longer than the optical pulse duration. 
To avoid distortion of the optical signal, we design the electrical pulse with length $\taus = 2\taumod + \tauop$, such that the shorter base of the trapezoidal modulation window is sufficiently wide to fully accommodate the optical pulse. 
Since $\taus > \tauop$, the Sagnac asymmetry $\taud$ can be reduced such that the shorter base of the trapezoid occurs immediately after the co-propagating component ends, allowing some overlap between the ramp of the trapezoid and the rectangular pulse, as long as the modulating region remains unaffected.
This implies $\taud-\taumod+2\taumod=\taus$ (see Fig. \ref{fig:phis_asymm}), which, when combined with the previous constraints, implies $\taud=\taumod+\tauop$. 
In this condition, when the optical pulse is placed fully within the shorter base of the trapezoid, the co-propagating modulation should be shifted by $\taus=2\taumod+\tauop$. 
Taking into account all components, a minimum repetition period of $\tsym=4\taumod+2\tauop$ is obtained. 
The resulting scheme is illustrated in Fig.~\ref{fig:modulations}(a).

\subsection{Differential modulation}

When using differential modulation, it is worth distinguishing between asymmetric and symmetric configurations. 
For the asymmetric configuration, since now the symbols do not require a temporal shift with respect to each other, the minimum symbol period becomes $\tsym=\taud + \taumod + \taus$, where there is no longer the constraint of $\taus\ge2\taumod+\tauop$, since only the co-propagating component is used for applying the modulation, but still keeping the constraint $\taud=\taus+\taumod$. 
To modulate the entire optical pulse equally, it is required that $\taus\ge\tauop$, thus obtaining $\tsym=2(\taumod+\taus)$.

On the other hand, when utilizing differential modulation with the Sagnac in a symmetric configuration, the minimum symbol period is reduced by $\taud$, obtaining instead $\tsym=\taumod+\taus$, once again with $\taus\ge\tauop$, which increases the maximum repetition rate by a factor of $2$ compared to the asymmetric case. 
However, this increase in repetition rate comes with the cost of increasing the required voltage by a gain factor of $g_V={{2 \taumod}/{(2 \taumod - \taus)}}$ as defined in Sec. \ref{sec:sagnac-symm}. 

\subsection{Balanced modulation}
When using balanced modulation, the maximum repetition rate is slightly reduced compared to the differential case in the symmetric configuration, since now the duration of the electrical signal $\taus$ is longer, requiring the use of two voltage pulses with inverted sign and equal duration $\taus/2$.
To guarantee that the entire optical pulse is modulated equally, $\taus/2\ge\tauop$.

For this case, the maximum repetition rate is defined as $\tsym=\taus+\taumod\ge2\tauop+\taumod$.
However, in comparison to the differential case, the use of balanced modulation relaxes the condition for higher voltage, making it potentially less costly to implement.
It is worth noting that this type of modulation also helps mitigate the patterning effect in amplifiers because the amplification depends on the average signal level, which in this case is zero and is not influenced by the sequence of consecutive pulses.

\section{Conclusion}
Due to their intrinsic phase stability and architectural flexibility, Sagnac-loop modulation schemes have been, and will continue to be, key components for quantum communications. This work provides a complete, general model for Sagnac modulators that generalizes existing schemes and offers a practical framework applicable to a broad class of loop modulators in the literature. Additionally, we presented and experimentally validated a symmetric configuration of the Sagnac loop with a differential modulation driving of the EOMs that enables stable operation and allows intensity and polarization modulation as well as the generation of optical pulses directly within the interferometer at higher repetition rates with respect to other traditionally used methods. 

Here, we demonstrated the use of the symmetric iPOGNAC at almost its theoretical maximum repetition rate of $R=1.5$~GHz (with the components used), achieving a polarization extinction ratio of $>23$~dB on both encoding bases. 
To provide further validation, we also show two modulation schemes for the symmetric polarization and intensity Sagnac modulator, and compare the measured output intensity with the one predicted by our model. 
When taking into account the non-idealities both on the driving electrical signal and optical detection system, the model is able to accurately predict the output of the Sagnac modulator. This combination of theory and practice provides a solid foundation for future photonic and quantum communication applications.

\section*{Acknowledgements}
M.R.B. acknowledges support from the European Union's Horizon Europe Framework Programme under Marie Sklodowska Curie Grant No. 101072637, Project Quantum-Safe Internet (QSI).
A.D.T. acknowledges the financial support of Concessioni Autostradali Venete (CAV) S.p.A. in the framework of the doctoral scholarship agreement 38° Ciclo between CAV and the University of Padova.
K.V. acknowledges support by grant No. 63132 from the John Templeton Foundation, as recipient of an Enrico Fermi Fellowship awarded through the Center for SpaceTime and Quantum. 
This work is partially supported by ICSC – Centro Nazionale di Ricerca in High Performance Computing, Big Data and Quantum Computing, funded by European Union – NextGenerationEU.

The opinions expressed in this publication are those of the author(s) and do not necessarily reflect the views of the respective funding body.

The content is object of Italian Patent Application No. 102025000008883 filed on 18.04.2025 \cite{patentSymmetriciPognac}.

\section*{Author Contributions}
The identification of limitations related to asymmetry of the Sagnac modulators and the initial studies to provide a solution for the problem were done  F.B., C.A., A.S., M.A., G.V., P.V..
The experimental design and implementation were carried out by F.B., M.R.B., A.D.T., C.A., M.A., and K.V..
C.A. and M.R.B. performed the mathematical validation and consistency checks.
F.B. and G.V. contributed to the conceptualization and model development realizing the transfer modulator function.
F.B. discovered the balanced modulation.
The concept of pulse generation was introduced by K.V. and M.A..
The manuscript was drafted and written by F.B., G.V., M.R.B., A.D.T., C.A., and K.V..
Critical revision and editing were performed by P.V., M.A., and A.S..

\bibliography{references}

\end{document}